# Impact of Top SiO$_2$ interlayer Thickness on Memory Window of Si Channel FeFET with TiN/SiO$_2$/Hf$_{0.5}$Zr$_{0.5}$O$_2$/SiO$_x$/Si (MIFIS) Gate Structure

Tao Hu, Xianzhou Shao, Mingkai Bai, Xinpei Jia, Saifei Dai, Xiaoqing Sun, Runhao Han, Jia Yang, Xiaoyu Ke, Fengbin Tian, Shuai Yang, Junshuai Chai, Hao Xu, Xiaolei Wang, Wenwu Wang, and Tianchun Ye

*Abstract*—We study the impact of top SiO$_2$ interlayer thickness on memory window of Si channel FeFET with TiN/SiO$_2$/Hf$_{0.5}$Zr$_{0.5}$O$_2$/SiO$_x$/Si (MIFIS) gate structure. The memory window increases with thicker top SiO$_2$. We realize the memory window of 6.3 V for 3.4 nm top SiO$_2$. Moreover, we find that the endurance characteristic degrades with increasing the initial memory window.

## I. Introduction

Hafnia(HfO$_2$) based silicon channel ferroelectric field-effect transistors (HfO$_2$ Si-FeFETs) have been extensively studied [1-18], as a strong candidate for non-volatile memory, thanks to the discovery of ferroelectricity in doped-HfO$_2$ [19]. The memory window (MW) of HfO$_2$ Si-FeFETs is generally required to be larger than 4 V to meet application in multi-bit memory cells. Recently, inserting a dielectric layer (e.g. SiO$_2$ or Al$_2$O$_3$) between the metal gate and ferroelectric layer was found to be an effective method to significantly improve the MW [20-25]. However, the impact of the top SiO$_2$ interlayer thickness on the MW is still unclear and this hinders the device design. Therefore, we report for the first time the dependence of the MW on the top SiO$_2$ interlayer thickness in this work. The MW increases with thicker SiO$_2$. We realize a maximum MW of 6.3 V by inserting a 3.4 nm SiO$_2$ top interlayer. Furthermore, we study the dependence of the endurance on initial MW. We find that the endurance characteristic degrades with increasing the initial MW. We realize an initial MW of 5.3 V by inserting a 3.4 nm SiO$_2$ top interlayer with an endurance of $7\times10^3$ cycles and a retention of 10 years.

This work was supported by the National Natural Science Foundation of China under Grant No. 92264104 and 52350195. (Corresponding author: Xiaolei Wang)
Tao Hu, Xianzhou Shao, Mingkai Bai, Xinpei Jia, Saifei Dai, Xiaoqing Sun, Runhao Han, Jia Yang, Xiaoyu Ke, Fengbin Tian, Shuai Yang, Junshuai Chai, Hao Xu, Xiaolei Wang, Wenwu Wang, and Tianchun Ye are with Institute of microelectronics of the Chinese Academy of Sciences, Beijing 100029, China. The authors are also with University of Chinese Academy of Sciences, Beijing 100049, China (e-mail: wangxiaolei@ime.ac.cn).

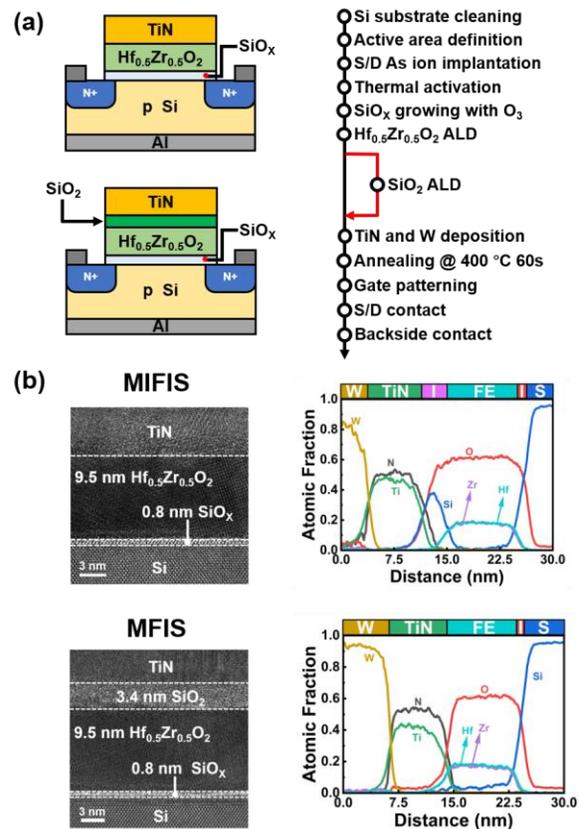

Fig. 1. (a) Schematic of the HfO$_2$ Si-FeFET device structure and fabrication process flow. (b) HRTEM images and EDS of the MIFIS and MFIS structures.

## II. Device Fabrication and Characterization

Fig. 1(a) shows the schematic of the devices and fabrication process flow. In this work, there are two different gate stacks. One is TiN/Hf$_{0.5}$Zr$_{0.5}$O$_2$/SiO$_x$/Si (MFIS) as the control sample. The other is TiN/SiO$_2$/Hf$_{0.5}$Zr$_{0.5}$O$_2$/SiO$_x$/Si (MIFIS) with a 0.85, 1.7, 2.55, or 3.4 nm SiO$_2$ top interlayer.

These devices are fabricated in an 8-inch P-type silicon wafer with a resistivity of 8-12 Ω·cm using a gate-last process. Firstly, the source and drain regions are formed by implanting

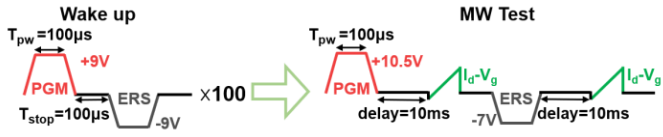

Fig. 2. Details of electrical measurement.

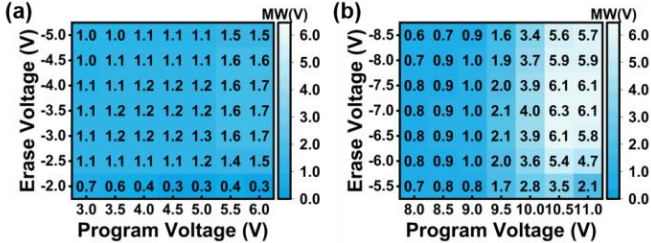

Fig. 3. The MW mapping of (a) MFIS and (b) MIFIS with 3.4 nm SiO₂ as a function of the pulse amplitude under the pulse width of 100 μs.

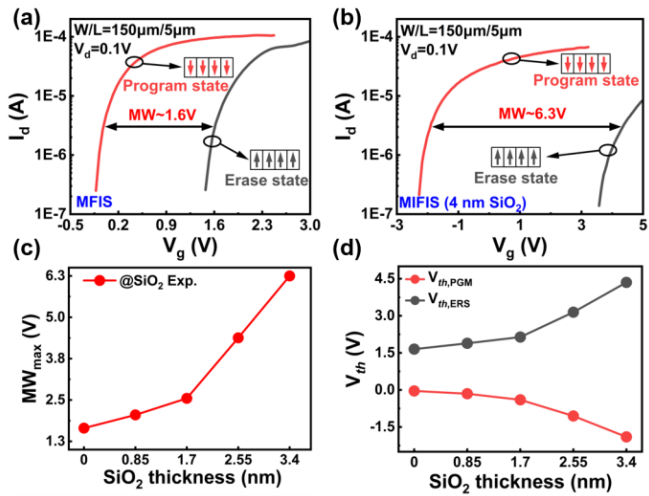

Fig. 4. (a) $I_d$–$V_g$ curves of maximum MW for (a) MFIS and (b) MIFIS with 3.4 nm SiO₂ top interlayer. The dependence of the (c) maximum MW and (d) $V_{th}$ on the top SiO₂ interlayer thickness.

As ions with an energy of 50 KeV and a dose of 2 × 10¹⁵ cm⁻². After that, these devices are annealed at 1050 °C for 5 s under the N₂ atmosphere for dopant activation. Subsequently, the gate stack was formed. The 0.8 nm SiOₓ botten interlayer was grown by O₃ oxidation at 300 °C. Then, 9.5 nm Hf₀.₅Zr₀.₅O₂ and 0.85-3.4 nm top interlayer SiO₂ were grown by atomic layer deposition (ALD) at 300 °C. After that, 10 nm of TiN and 75 nm W were grown by sputtering. The devices are annealed at 400 °C under an N₂ atmosphere for 60 s to form the orthorhombic phase. Finally, the forming gas annealing (FGA) was performed at 450 °C in 5%-H₂/95%-N₂. The gate width/length (L/W) of these devices is 5/150 μm. The electrical measures were performed by Keysight B1500A. The threshold voltage ($V_{th}$) is defined by the constant current method.

Fig. 1(b) shows High-Resolution Transmission Electron Microscopy (HRTEM) images and Energy Dispersion Spectrometer (EDS) results for both MIFIS and MFIS structures. For the MIFIS structure, the presence of a peak concentration of Si at the TiN/Hf₀.₅Zr₀.₅O₂ interface confirms the presence of the SiO₂ layer.

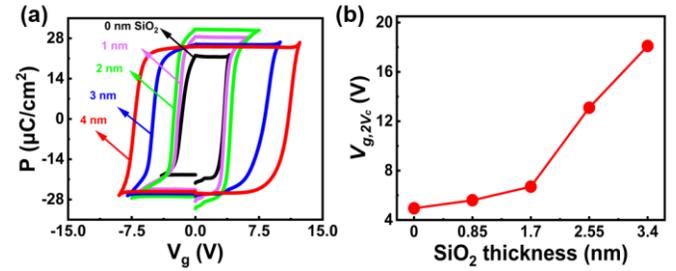

Fig. 5. (a) The results of the PUND measurements. (b) The dependence of $V_{g,2Vc}$ on the top SiO₂ thickness.

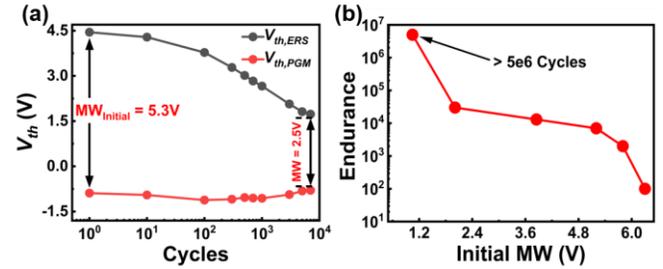

Fig. 6. (a) $V_{th}$ as a function of cycles for the MIFIS structure with 3.4 nm top SiO₂. (b) The dependence of endurance on initial MW for MIFIS structure.

## III. RESULTS AND DISCUSSIONS

We investigate the dependence of the MW on pulse Voltage amplitude. Fig. 2 shows the measurement waveforms. The pulse width is set as 100 μs. Fig. 3 shows the MW mapping result for the MFIS and MIFIS structure with 3.4 nm SiO₂ (unless specified otherwise in the following context, MIFIS structure refers to the MIFIS structure with the 3.4 nm SiO₂). The maximum MW is 6.3 V for the MIFIS structure device, while the maximum MW is 1.7 V for the MFIS sample. Fig. 4(a) and (b) show the $I_d$-$V_g$ curves corresponding to the maximum MW for the MFIS and MIFIS structures. The physical origin of the larger MW compared with the control sample is due to the charge trapping at the top SiO₂/Hf₀.₅Zr₀.₅O₂ interface.

We study the impact of top SiO₂ thickness on the maximum MW. Fig. 4(c) shows the impact of top SiO₂ thickness on the maximum MW. The maximum MW increases with thicker top SiO₂ thickness. Fig. 4(d) shows the dependence of the threshold voltage ($V_{th}$) on the top SiO₂ thickness. Fig. 5(a) shows the PUND measurement results. The bulk, source, and drain terminals were shorted during the measurement. The polarization vs gate voltage ($P$-$V_g$) curve broadens with increasing top SiO₂ thickness. We define the distance between the intersections of the $P$-$V_g$ curve and the $V_g$ axis as $V_{g,2Vc}$. Fig. 5(b) shows the dependence of $V_{g,2Vc}$ on the top SiO₂ thickness.

We conduct the endurance measurement for the MIFIS structure. Fig. 6(a) shows the endurance characteristics of MIFIS structures. The MIFIS sample shows endurance of 7 × 10³ cycles when the initial MW is 5.3 V. Fig. 6(b) shows the dependence of endurance on the initial MW. The endurance degrades with a larger initial MW.

We study the retention characteristics. Fig. 7 shows the retention characteristics for the two structures. We apply a preset waveform to put the device into the same state before

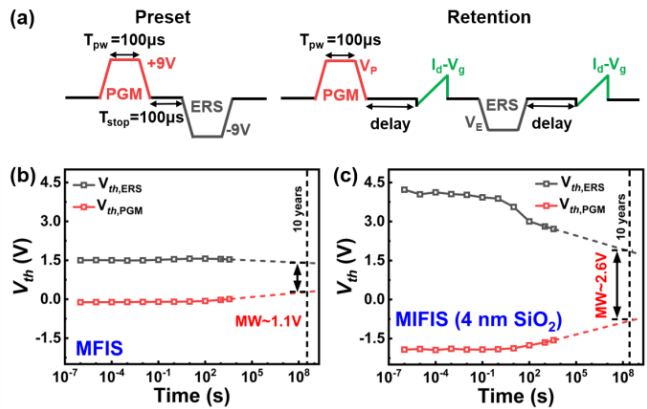

Fig. 7. (a) The measurement waveform of retention. Retention characteristics measured at room temperature of (b) the MFIS structure and (c) the MIFIS structure with 3.4 nm top SiO$_2$.

TABLE I
Comparision of recent research with our work

| | Structure | Top IL | MW | Endurance | Retention |
|---|---|---|---|---|---|
| Lee et al. from SK Hynix [21] (2022) | MIFIS | SiO$_2$ | 3.8 V | unkonw | 10 years |
| Das et al. from GIT [24] (2023) | MFIFIS | Al$_2$O$_3$ | 7.1 V | unkown | 10 years |
| Suzuki et al. from Kioxia [23] (2023) | MIFIS | unkonw | 2.3 V | ~10$^6$ cycles | 10 years |
| Lim et al. from Samsung [22] (2023) | MIFIS | unkonw | 3.1 V | ~10$^6$ cycles | 10 years |
| Yoon et al. from SK Hynix [20] (2023) | MIFIS | unkonw | 5.04 V | ~3×10$^3$ cycles | 10 years |
| Hu et al. from Imecas [25] (2024) | MIFIS | Al$_2$O$_3$ | 4.1 V | ~1×10$^4$ cycles | 10 years |
| This work | MIFIS | SiO$_2$ | 5.8 V | ~3×10$^3$ cycles | 10 years |

each retention measurement. Both devices have a retention lifetime beyond 10 years.

Table I shows the benchmark of our work. Our work shows a larger MW (5.8 V) when the endurance is ~3000 cycles and the retention can achieve 10 years.

## IV. CONCLUSIONS

In this work, we study the impact of the top SiO$_2$ thickness on the MW. We find that the MW increases with thicker top SiO$_2$ interlayer. Furthermore, we also investigate the dependence of the endurance on initial MW, and find that the endurance characteristic degrades with increasing the initial MW. By inserting a 3.4 nm top SiO$_2$ interlayer between the gate metal TiN and the ferroelectric Hf$_{0.5}$Zr$_{0.5}$O$_2$, we achieve the MW of 6.3 V with an endurance of 200 cycles. When the MW is ~5.8 V, the endurance can achieve 3000 cycles and the retention is larger than 10 years.